\documentclass[doublecol]{epl2} 

\usepackage{graphicx}
\usepackage{dcolumn}
\usepackage{amssymb}
\usepackage{bm}
\usepackage{epsfig}
\usepackage{psfrag}
\usepackage{setspace}

\def\3dots{\:\raisebox{-0.5ex}{$\stackrel{\textstyle.}{:}$}\:}
\def\beq{\begin{equation}}
\def\eeq{\end{equation}}
\def\bea{\begin{eqnarray}}
\def\eea{\end{eqnarray}}

\title{Out-of-equilibrium microrheology using optical tweezers to probe directional viscoelastic properties under shear}
\shorttitle{Out-of-equilibrium microrheology using optical tweezers}

\author{Manas Khan and A. K. Sood}
\shortauthor{M. Khan \etal}
\institute{Department of Physics, Indian Institute of Science, Bangalore - 560012, India                    
}

\pacs{87.80.Cc}{Optical trapping}
\pacs{83.60.Fg}{Shear rate dependent viscosity}
\pacs{87.15.N-}{Properties of solutions of macromolecules}
\pacs{87.19.rh}{Fluid transport and rheology}

\abstract{Many wormlike micellar systems exhibit appreciable shear thinning due to shear induced alignment. As the micelles get aligned introducing directionality in the system, the viscoelastic properties are no longer expected to be isotropic. An optical tweezers based active microrheology technique enables us to probe the out-of-equilibrium rheological properties of a wormlike micellar system simultaneously along two orthogonal directions - parallel to the applied shear, as well as perpendicular to it. While the displacements of a trapped bead - in response to active drag force carry signature of conventional shear thinning, its spontaneous position fluctuations along the perpendicular direction manifest an orthogonal shear thickening, an effect hitherto unobserved.}

\begin{document}

\maketitle


Viscoelastic fluids are distinguished in showcasing novel flow properties \cite{larson}. In most of the cases, such interesting rheological behaviors originate from the modulation of interaction between the structural units induced by the flow perturbations, namely shear stress and shear strain. To be more specific, in wormlike micellar solutions all the flow properties are governed by the length, entanglement and relative orientation of the micelles. In these systems, the non-Newtonian behaviors characterized by a strain-rate-dependent effective viscosity arise from a flow-induced alignment of the micelles \cite{sans, sals, electron, gemini}. SANS \cite{sans}, SALS \cite{sals} patterns and electron micrographs \cite{electron} provide direct evidence of this orientational ordering along the direction of applied shear. It poses a very important question - whether this structural ordering in such sheared systems imposes an anisotropy in their flow properties. While the micellar alignment facilitates shear thinning along the direction of flow, it may alter the rheological parameters differently along the other directions. Unfortunately the conventional rheology experiments, that measure a system's response parallel to the applied shear, fail to probe this shear-induced anisotropy thereby leaving this important issue unaddressed.

In this Letter we report an experimental probe to study the anisotropy in rheological properties of a sheared viscoelastic medium. Using an optical tweezers based active microrheology technique \cite{dragcolloid, dragcolloid2, drag} we have measured the flow properties of a wormlike micellar system simultaneously along two orthogonal directions; the direction of applied shear, and more importantly, the one perpendicular to it. A micron sized polystyrene bead held in a tight optical trap is dragged through the medium at a constant velocity in order to shear the system. While this active forcing drives the medium to a defined sheared state along the direction of drag, the spontaneous thermal fluctuations of the trapped bead apply passive thermal forcing along an orthogonal direction. Taking advantage of the Fluctuation Dissipation Theorem (FDT) \cite{kubo}, we thereby probe the system's loss modulus through the bead's position fluctuations along the direction normal to that of the applied shear. A strong optical trap ensures the bead's driven motion, relative to the medium, to be strictly one dimensional along the direction of the drag. Thus the system can be pushed beyond linear response along one direction while the surrounding fluid still faithfully behaves like an equilibrium system by virtue of the one dimensional motion of the microbead. This approach permits us to utilize the FDT in investigating the viscoelastic properties of a CTAT (cetyltrimethylammonium tosylate) system along the perpendicular direction to the applied shear. Moreover, the shear thinning along the parallel direction can simultaneously be examined by measuring the drag force on the trapped bead. To the best of our knowledge, the measurement of rheological properties along a direction orthogonal to that of the applied shear has not been reported earlier.

\begin{figure}[htbp]
\includegraphics[width=0.4\textwidth]{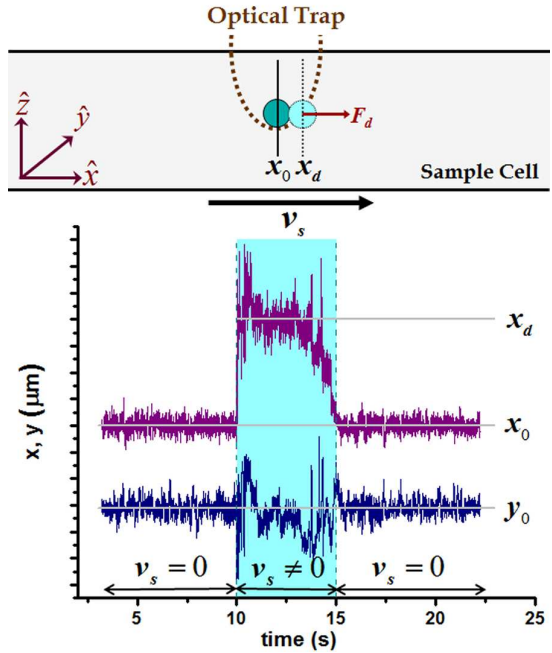}\\
\caption{(Color online) The schematic of the experiment. The sample stage is given a velocity $v_{s}$ to apply a shear stress on the medium by the trapped bead along the direction of relative velocity ($\hat{x}$). In the course of this motion the drag force ($\vec{F_{d}}$) pushes the trapped bead to a new equilibrium position $x_{d}$ from its otherwise mean position $x_{0}$. Along $\hat{y}$, its spontaneous position fluctuation is always about the same mean value $y_{0}$.} \label{expt}
\end{figure}

For this active microrheology experiment, a very dilute suspension of 1.9 $\mu m$ polystyrene beads in 1 wt\% CTAT solution is loaded in a sample cell made up of two cover glasses separated and sealed by a 125 $\mu m$ thick double stick tape. A polystyrene bead is optically trapped at a height of $\sim$ 25 $\mu m$ from the bottom plate by a tightly focused infrared (1064 $nm$) laser beam. Another 680 $nm$ laser beam, collinear to the trapping beam, is used to image the trapped bead on a quadrant photo diode (QPD) for its position detection. The QPD current signals, captured at 1 $kHz$ bandwidth, is converted and amplified to voltage signals which linearly correspond to the $x$ and $y$ positions of the microbead. The position fluctuations of the probe bead are then analyzed to obtain the microrheological properties of the medium. In order to drive the system to a sheared state, the sample stage is given a constant velocity $v_{s} \hat{x}$ while the microbead is held stationary in the moving medium by the static optical trap (Fig \ref{expt}). The stage motion is alternately kept on and off for a time interval of $\Delta t_{v_{s} \neq 0}$ and $\Delta t_{v_{s}=0}$ respectively and every $v_{s} \hat{x}$ movement is retraced by a $-v_{s} \hat{x}$ motion. Fluctuation signals are recorded continuously in this process. As $v_{s}$ is increased, to set a higher strain-rate, $\Delta t_{v_{s} \neq 0}$ is decreased, keeping the travel length under a limiting value. The experiment is repeated for four different shear rates corresponding to the stage speeds ($v_{s}$) 10 $\mu m/s$, 20 $\mu m/s$, 30 $\mu m/s$ and 40 $\mu m/s$. The bulk rheological properties of the same sample are measured using a Paar Physica MCR 300 rheometer.

Since this experiment demands a very tight optical trap to achieve a strictly one dimensional (along $\hat{x}$) constant velocity drive mode, a high power infrared laser beam is used. The laser power delivered to the sample cell, as measured after the objective, is $\sim$ 150 $mW$. This intense beam creates a notably strong optical trap holding the polystyrene bead tightly. The trap stiffness in this viscoelastic medium cannot be measured in a direct fashion. However, it is extracted as a part of the microrheology data analysis (discussed in a later section) and takes the values: $\kappa_{x}=$ 28.6 $pN/\mu m$, $\kappa_{y}=$ 27.5 $pN/\mu m$. At this high laser power, local heating poses a major concern. It might alter the local rheological properties of the sample or affect surface smoothness of the probe bead, eventually leading to erroneous measurements. To tackle this crucial issue, the trap position in the sample cell as well as the probe bead is changed after every 60 $s$ which is the time span of a single data set. The sample cell is also replaced with a new one after recording every 10 data sets to maintain the sample at room temperature.

The position fluctuation data captured in the time interval $\Delta t_{v_{s}=0}$ (when the trapped bead is not actively perturbing the system) provide the system's microrheological properties through the FDT. This theorem enables us to derive the response function of the system $\chi (f) = \chi'(f)+i\chi''(f)$ directly from the equilibrium fluctuation data $x(t)$ and $y(t)$ through the following equation \cite{th1, th2}
\begin{equation}
\chi''(f) = \frac{\pi}{2 k_{B} T} f S(f),
\label{eqfdt}
\end{equation}
where $k_{B}$ is the Boltzmann constant, $T$ is the system temperature and $S(f)$ is the single-sided Power Spectral Density (PSD) of the time series $x(t)$ (or $y(t)$). A Kramers-Kronig relation can then be used to evaluate the real part of the response function $\chi'(f)$ as
\begin{equation}
\chi'(f) = \frac{2}{\pi}\int^{\infty}_{0} dt \cos(ft) \int^{\infty}_{0} d\xi \sin(\xi t) \chi''(\xi).
\label{kkt}
\end{equation}
The optical trap response, being elastic in nature, gets added to the system's elastic response thereby giving an overestimate of the actual system response. To subtract the trap effect, the response function needs to be corrected as
\begin{equation}
\alpha (f) = \frac{\chi (f)}{1-\kappa \chi (f)},
\end{equation} 
where $\alpha$ is the corrected system response function and $\kappa$ is the trap stiffness. The corrected response function is then converted to the complex shear modulus of the system $G(f) = G'(f)+iG''(f)$ by the relation,
\begin{equation}
G(f) = \frac{1}{6 \pi a \alpha (f)},
\label{g12}
\end{equation}
$a$ being the radius of the bead. $G'(f)$ and $G''(f)$ are the elastic and loss moduli, respectively. Eq. \ref{g12} can be simplified to write in terms of $\chi'(f)$ and $\chi''(f)$ as
\begin{eqnarray}
G'(f) = \frac{1}{6 \pi a}\frac{\chi'}{\chi'^{2}+\chi''^{2}} - \frac{\kappa}{6 \pi a};\\
G''(f) = \frac{1}{6 \pi a}\frac{\chi''}{\chi'^{2}+\chi''^{2}}.
\label{gg}
\end{eqnarray}
Since measuring the trap stiffness in a viscoelastic medium is not so straightforward, it is customary to consider the average value, calculated over first few points in the flat region of the uncorrected $G'(f)$ as the correction factor for the trap effect \cite{th1}. The correction factor, which is equal to $\frac{\kappa}{6 \pi a}$, can then be safely used to calculate the true trap stiffness $\kappa$ in the medium.

The state of the system is dramatically changed when (in time interval $\Delta t_{v_{s} \neq 0}$) it is actively sheared along $X$ by the trapped bead that is dragged through the medium with a velocity $v_{s}\hat{x}$ (Fig \ref{expt}). Neglecting the elastic response of the medium, the drag force $F_{d}$ acting on the bead is roughly calculated as $F_{d} = 6 \pi \eta_{x} a v_{s}$, where $\eta_{x}$ denotes the viscosity of the medium along $X$. This drag force, $F_{d}$, pushes the trapped bead from its otherwise mean position $x_{0}$ to a new equilibrium position $x_{d}$, where the restoring force balances the drag force \cite{drag}, i.e. 
\begin{equation}
F_{d} = 6 \pi \eta_{x} a v_{s} = \kappa_{x} (x_{d} - x_{0}),
\end{equation} 
$\kappa_{x}$ being the force constant of the trap along $X$. Different stage velocities $v_{s}$, which can be represented as shear strain rates $\dot{\gamma}_{x}$, would result in varying shear viscosities $\eta_{x}$. Therefore, along the direction of active drive, $\eta_{x} (\dot{\gamma}_{x})$ can be evaluated as 
\beq
\eta_{x} = \kappa_{x} (x_{d} - x_{0})/6 \pi a v_{s}.
\label{eta}
\eeq               
It should be noted that neglecting the elastic response of the medium would lead to an underestimation of $\eta_{x}$. The stage velocities ($v_{s}$) can be converted to equivalent strain rates ($\dot{\gamma}_{x}$) using the relation \cite{gamma}
\beq
\dot{\gamma}_{x} = \frac{\sigma_{avg}}{\eta} = \frac{6 \pi \eta a v_{s}}{\pi a^{2}}.\frac{1}{\eta} = \frac{6 v_{s}}{a},
\label{rate}
\eeq
where $\sigma_{avg} = \frac{6 \pi \eta a v_{s}}{\pi a^{2}} $ is the average uniform stress applied on the fluid by the microbead.

In course of the translational motion of the stage, the viscoelastic medium (CTAT) flows past the trapped bead at a velocity $v_{s}$. The strong optical trap holds the polystyrene bead rigidly enough so that the microstructures in the medium cannot impose any velocity fluctuations to the bead while passing by. Therefore, the velocity of the trapped bead, relative to the fluid, will not change its value or direction in course of this motion. This permits us to treat the drive along $X$ as a constant strain rate process. More importantly, this ensures that there is no driven movement of the bead along $Y$. Consequently, the bead's position fluctuations along $Y$ can be treated as the spontaneous thermal fluctuations connected to the dissipative property of the medium (along $Y$) through FDT. It is noteworthy that this position fluctuations along $Y$, $y(t)_{\dot{\gamma}_{x} \neq 0}$, are not the same as the unperturbed equilibrium fluctuations $y(t)_{\dot{\gamma_{x}} = 0}$. The active shear along $X$ induces a micellar organization in the surrounding fluid which, in turn, regulate the fluctuation properties of the bead along $Y$. The effect of this structural organization can, therefore, be probed by analyzing $y(t)_{\dot{\gamma}_{x} \neq 0}$. Using FDT and the subsequent recipe (Eq. \ref{eqfdt} to Eq. \ref{gg}), as explained above, the storage and loss modulus along $Y$ can be evaluated for different strain rates along $X$. We denote these viscoelastic moduli as $G'_{y}(\dot{\gamma}_{x} \neq 0)$ and $G''_{y}(\dot{\gamma}_{x} \neq 0)$ respectively.

\begin{figure}[htb]
\includegraphics[width=0.48\textwidth]{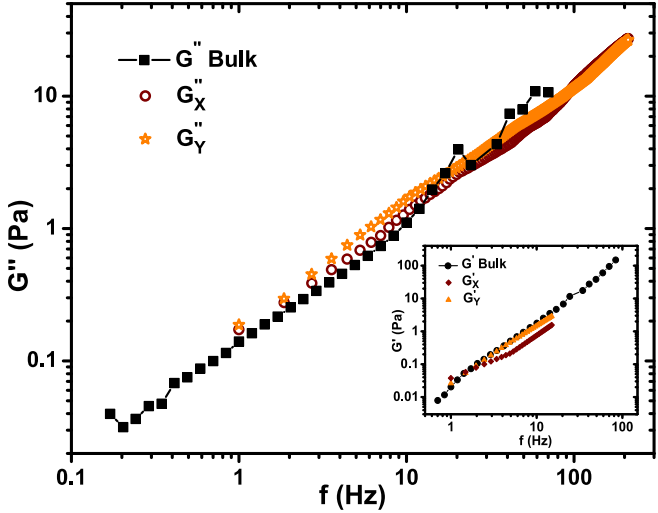}\\
\caption{(Color online) The microrheological loss moduli, $G''_{x}$ and $G''_{y}$, computed from the equilibrium fluctuation data $x(t)$ and $y(t)$ respectively, are compared with the bulk rheology data. Corresponding microrheological storage moduli (corrected), $G'_{x}$ and $G'_{y}$, are shown alongside the bulk $G'$ data in the Inset.}
\label{eqbr}
\end{figure}

In this experiment, for each $v_{s}$ values, the system is alternately pushed to a driven state (for time $\Delta t_{v_{s} \neq 0}$) and pulled back to the initial equilibrium state (for time $\Delta t_{v_{s} = 0}$) for  several cycles and the position fluctuation data ($x(t)$, $y(t)$) of the trapped bead are recorded continuously. The zero-shear datasets captured in this manner are analyzed to ensure that the starting point of each driven states are indeed the unimpaired equilibrium state of the system and the data captured in the sheared states are free from any artifacts caused by local heating. The average storage ($G'$) and loss moduli ($G''$) obtained from the equilibrium fluctuation datasets are plotted and compared with the bulk rheological properties of the sample in Fig. \ref{eqbr}. A very good match between the equilibrium viscoelastic moduli computed from $x(t)$ ($G'_{x}$ and $G''_{x}$) and $y(t)$ ($G'_{y}$ and $G''_{y}$) clearly confirms that prior to each sheared states, the system behaves purely as an isotropic medium. In addition, it can also be concluded that the rheological properties of the system as measured in this experiment, even in the in-between driven states, reflect the true behavior of the medium and are not altered by any unwanted effects that might arise while working with high laser powers. The trap stiffness $\kappa$, both along $X$ and $Y$, in the CTAT system are also evaluated from the correction factor ($\frac{\kappa}{6 \pi a}$) of the corresponding storage moduli and used in other calculations.

A slight deviation of the microrheological data from the corresponding bulk values can be attributed to the inherent imperfection of one-point-microrheology. The tracer bead, that conveys the medium's flow properties through its equilibrium fluctuations, gets coupled to the bulk medium by the immediate surrounding whose rheological properties are modified by the introduction of the probe particle \cite{shell, shell2}. The viscoelastic moduli measured in this way actually represent some convolution of the perturbed layer (shell) and the bulk material properties.

\begin{figure}[htbp]
\includegraphics[width=0.4\textwidth]{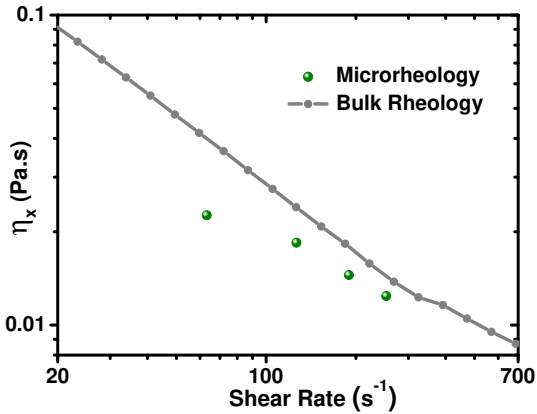}\\
\caption{(Color online) The shear thinning along $X$, $\eta_{X}(\dot{\gamma}_{x})$, as observed in the active microrheology (along $X$) are compared with the bulk flow curve.}
\label{etax}
\end{figure}

In the driven states, the trapped bead, moving at a constant speed ($v_{s}$) with respect to the medium, actively shears the surrounding system. In this process, due to the drag force $F_{d}$, the bead gets displaced from its otherwise mean position in the optical trap. Measuring this displacement, ($x_{d} - x_{0}$), the shear viscosity of the medium along $X$ ($\eta_{x}$) is obtained for different strain rates $\dot{\gamma}_{x}$ using Eq. \ref{eta} and Eq. \ref{rate}. The $\eta_{x}(\dot{\gamma}_{x})$ values are plotted and compared to the bulk flow curve in Fig. \ref{etax}. As this measurement does not include the elastic response of the medium, it underestimates the shear viscosity to some degree. The deviation is more prominent at the lower shear rate values.

\begin{figure}[htbp]
\includegraphics[width=0.48\textwidth]{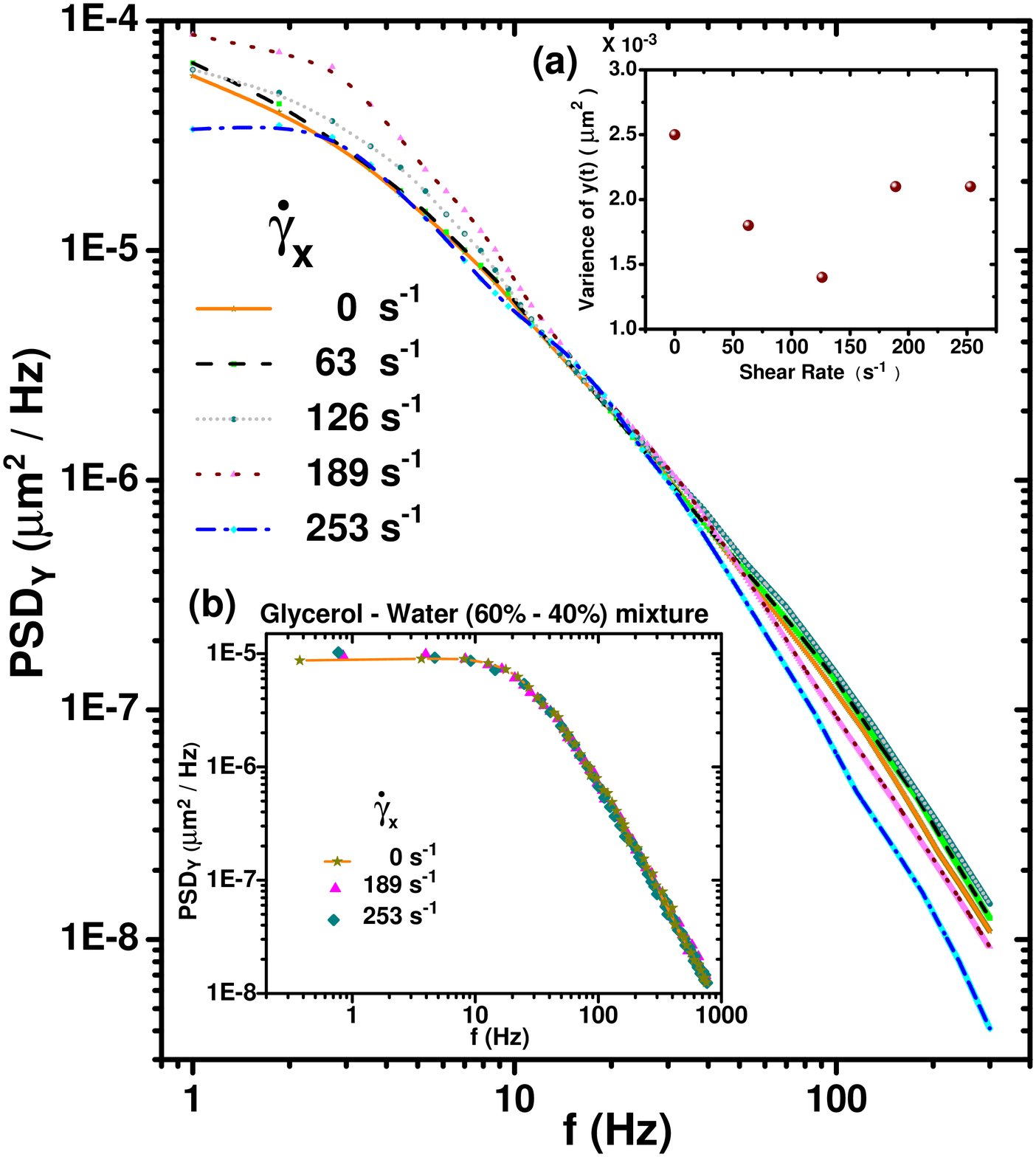}\\
\caption{(Color online) Power Spectral Densities ($PSD_{y}$) of the spontaneous position fluctuation of the microbead along $Y$ are shown for different shear rates $\dot{\gamma}_{x}$. Compared to the equilibrium data, initially the PSDs go up slightly for the first two shear rates and then decrease significantly as $\dot{\gamma}_{x}$ is increased further. Inset(a) shows the variance of the position fluctuations along $Y$ ($y(t)$) plotted against shear rates. Inset(b) exhibits the results of identical experiments performed on a Newtonian fluid (glycerol-water 60\%-40\% mixture) using exactly the same settings. In this case, even for the highest two $\dot{\gamma}_{x}$ values, the $PSD_{y}$ do not show any deviation from the equilibrium ($\dot{\gamma}_{x}=0$) data.}
\label{psd}
\end{figure}

While the bead applies active shear stress on the CTAT system along the $X$ direction thereby forcing a shear induced alignment, which is manifested by the shear thinning along $X$, its spontaneous position fluctuation along $Y$ remains unaffected by the drive. In other words, the $Y$-position fluctuations of the bead do not comprise of any driven component. To support this proposition, the variance of $y(t)$ has been calculated and plotted against each shear rate values (Fig. \ref{psd}-Inset(a)). A non-monotonic variation of the variance, $Var(y(t))$, clearly establishes that $y(t)$ is not directly regulated by any component of the active shear strain along $X$. It also confirms that the bead's $Y$-position fluctuation data are free from any artifacts that may originate because of the high shear rates $\dot{\gamma}_{x}$.

Therefore, the system's response to the thermal shear along $Y$ can be analyzed (using Eq. \ref{eqfdt} to Eq. \ref{gg}) to obtain its modified viscoelastic properties in that direction for different shear rates along $X$. The power spectral densities (PSD) of $y(t)$ at varying $\dot{\gamma}_{x}$ are shown in Fig. \ref{psd}. In the PSD plot, the higher frequency regime reveals the system's dissipative properties whereas the lower frequency data represent the trap characteristics combined with the elastic response of the medium. In this figure, after a certain frequency, $\sim$ 30 $Hz$, the PSD curves start splitting up systematically and exhibit the system characteristics. The higher frequency fluctuations get enhanced slightly, compared to the unperturbed data, for the lowest two shear rates ($\dot{\gamma}_{x}$). However, the trend becomes completely opposite as the shear rate is increased further, where the fluctuations at higher frequencies diminish significantly for each increment in the shear rate. It is worth mentioning that when this experiment is repeated on a Newtonian fluid (glycerol-water 60\%-40\% mixture) for a consistency check, the PSDs of $y(t)$ at different $\dot{\gamma}_{x}$ do not show any variation from the equilibrium PSD (as shown in Fig \ref{psd}Inset(b)) and reflect the same bulk flow properties, as expected.

\begin{figure}[htbp]
\includegraphics[width=0.48\textwidth]{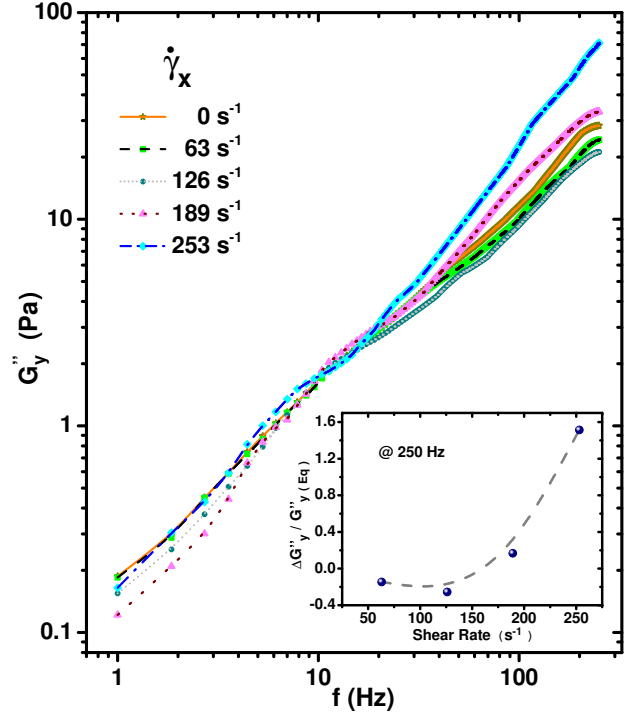}\\
\caption{(Color online) Loss moduli along $Y$, $G''_{y}$, are shown for different $\dot{\gamma}_{x}$, along with the unperturbed data set. In the Inset, the normalized increments of the loss modulus ($\frac{G''_{y}(\dot{\gamma}_{x})-G''_{y}(\dot{\gamma}_{x}=0)}{G''_{y}(\dot{\gamma}_{x}=0)}$) at a frequency value 250 $Hz$ have been plotted against the shear rate, $\dot{\gamma}_{x}$. The dashed line is a guide to the eye.}  
\label{gy}
\end{figure}

The PSD of $y(t)$ in various driven states have been analyzed to evaluate the loss modulus along $Y$. $G''_{y}$ for different driven states, including the equilibrium data set, have been displayed in Fig. \ref{gy}. In this figure, the featureless lower frequency side does not faithfully represent the system properties. It is noteworthy that to extract the system properties at the lower frequencies, an accurate calculation of the $\chi'(f)$ (Eq. \ref{kkt}) is essential. However, for a limited frequency dataset the Kramers-Kronig transformation along with the subsequent treatments does not produce consistent results. Nevertheless all the important system properties are distinctly visible in the higher frequency regime. For lower strain rates, the $G''_{y}$ decreases a bit from its unperturbed value. This effect can be attributed to the property of the shell which isolates the tracer bead from the bulk medium \cite{shell, shell2}. As the bead is dragged through the medium, the shell tries to follow the bead. In this process the shell needs to relax and reform itself at the newly moved position of the bead. Therefore, with increasing strain rates ($\dot{\gamma}_{x}$), the thickness and effect of the shell diminish and the viscoelastic properties of the bulk medium prevail. This shell-effect is manifested here, suppressing the medium properties, for the first two strain rates ($\dot{\gamma}_{x}$ = 63$s^{-1}$ and 126$s^{-1}$). At higher strain rates, when the shell can no longer follow the tracer bead, the bead gets exposed to the medium and can probe the true viscoelastic properties. At this regime (from $\dot{\gamma}_{x}$ = 126$s^{-1}$ to 253$s^{-1}$) the $G''_{y}$ increases significantly with the shear rate along $X$. To demonstrate it more clearly, the normalized increment of the $G''_{y}$, defined as $\frac{G''(\dot{\gamma}_{x})-G''(\dot{\gamma}_{x}=0)}{G''(\dot{\gamma}_{x}=0)}$, at a constant frequency 250 $Hz$, is plotted against the strain rates ($\dot{\gamma}_{x}$) in Fig. \ref{gy}-Inset. The loss modulus along $Y$ increases almost two folds as the strain rate along $X$ changes from 126 $s^{-1}$ to 253 $s^{-1}$.

Studying the passive and active microrheology (Fig. \ref{etax}) concurrently, it can be inferred that the enhanced loss modulus along $Y$ with increasing strain rates $\dot{\gamma}_{x}$, is the consequence of the same shear induced micellar alignment that causes a shear thinning along $X$. This experimental study clearly manifests that the shear induced reorganizations of the micro-structures in a viscoelastic medium not only change the viscosity along the direction of shear, but modify the system entirely and thereby regulate the flow properties accordingly along the orthogonal directions too.

To conclude, we have proposed and demonstrated a laser tweezers based microrheology technique to probe the viscoelastic properties of a wormlike micellar solution along a direction orthogonal to that of the applied shear. While active shear stress is applied along $X$, we exploit the spontaneous thermal stress along $Y$ to perform the active and passive microrheological studies simultaneously along the two perpendicular directions. Our results clearly show that the applied shear induces an overall anisotropy in the system. As the system shows reduced viscosity along the direction of applied shear, viscous modulus increases along the normal direction with increasing strain rates. This firmly suggests that under shear, the viscoelastic properties are not only a function of the strain-rate, but also dependent on the direction along which those are being measured. Therefore, for a more accurate representation, the shear viscosity of a viscoelastic system needs to be represented as $\eta_{\theta}$, where $\theta$ is the angle between the direction of applied shear and the direction along which the rheological properties are measured. For a shear thinning system, as in our case, $\eta_{\theta}$ takes its lowest value for $\theta = 0$ and attains the maximum at $\theta = \pi /2$. While the former phenomenon is well known as shear thinning, the latter could be termed as `orthogonal shear thickening'. This study, we hope, will motivate quantitative theoretical calculations to understand the rheological properties of such structured systems in a more complete fashion.

\acknowledgments

We thank Council of Scientific and Industrial research (CSIR), India, for financial support under Bhatnagar Fellowship.

\end{document}